# Comparison of Diffusive Motion of C60 on Graphene and Boron Nitride Surfaces

M. Vaezi [a], H.Nejat Pishkenari [b]*

[a] Institute for Nanoscience and Nanotechnology (INST), Sharif University of Technology, Tehran, Iran

[b] Mechanical Engineering Department, Sharif University of Technology, Tehran, Iran
*nejat@sharif.edu

**Abstract:** Inspired by macroscale vehicles, many molecular machines have been synthesized yet. Due to the importance of atomic manipulation at the surface, nanocars are of particular interest. Nanocar wheels play a crucial role in their motion. One of the popular candidate for using as a nanocar wheel is C60 molecule. In addition, the investigation of nanocar motion on two dimensional substrates is important due to formation of ripples on surface. Therefore, in this work we study the motion of C60 on graphene and boron nitride monolayer substrates at 300 K. The simulations are conducted by the means of molecular dynamics method. The trajectory of motion shows more displacement range when fullerene is moving on the graphene surface and as a result, it has a greater diffusion coefficient in this case. It has been showed that C60 motion on graphene and boron nitride obey a same super-diffusion regime at 300 K with different anomaly parameters.

## Introduction

The importance of atomic-scale manipulation has led scientist to fabricate machines at molecular level. Many types of molecular machines have been synthesized so far, such as molecular switches, molecular rotors and nanocars. Inspired by macroscale vehicles, nanocars have been created equipped with chassis, axles, and wheels. In previous investigations, it has been shown that, the wheels play a crucial role in the motion of nanocars, and their motion is significantly affected by the motion of wheels.[1] Among different type of wheels which have been utilized in the synthesis of nanocars, fullerene is a popular molecule.

In addition to the importance of nanocar wheels, substrates are also of particular importance in the motion of the nanocars. Concerning that 2D materials are able to form surface ripples and have a wavy morphology, it is remarkable to investigate the motion of wheels on them. The previous study characterized the influence of the ripples of graphene monolayer and double layers on the motion of C60 [2].

Here, we study the motion of fullerene on two different mono-layer substrates including graphene and boron nitride structures. We intent to compare the diffusive motion of C60 on these two substrates. The graphene and boron-nitride substrates are similar in lattice structure. Trajectory, diffusion coefficient and diffusion anomaly parameter of C60 molecule were computed by means of classical molecular dynamics (MD) simulations.

## Computational Methods

Molecular dynamics (MD) simulations have been utilized in order to study the motion of fullerene molecule on graphene and boron nitride monolayers. In both systems all internal vibrations of fullerene molecule are enabled.

The substrates are considered two square sheets with 12nm×12nm dimensions. Graphene substrate consists of 5684 carbon atoms, while boron nitride includes 5376 boron and nitrogen atoms. C60 centre of mass is placed at the centre of monolayers in lateral dimensions and vertically it is located near the equilibrium distance (6.4 A°). In order to have an unlimited motion for C60 molecule, periodic boundary condition is applied in X and Y directions. Fig. 1 illustrates the configuration of C60-graphene and C60-BN systems.

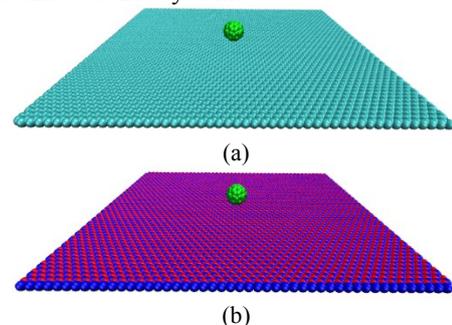

Fig. 1. Side view representation of (a) C60-graphene and (b) C60-BN systems

Classical molecular dynamic simulations were performed with the Large-scale Atomic/Molecular Massively Parallel Simulator (LAMMPS). Tersoff potential is used to apply intra layer interactions for boron nitride and graphene sheets. To describe the interactions of fullerene atoms, airebo potential is utilized. The potential between C60 and substrates is calculated using 6-12 Lennard-Jones potential with cut-off radius of 12 Å. Table 1 shows the LJ parameters that have been employed.[3]

Two Nose-Hoover thermostats with inertia parameters of 100 fs are used to control the temperature of C60 and substrate in each system. Both simulations are conducted at 300 K temperature. Both MD simulations are computed for 8 ns with 1 fs time step. The velocity Verlet integration method has been utilized.

Table 1. Lennard-Jones potential parameters used in simulations

|  | $\varepsilon$ (meV) | $\sigma$ (Å) |
|---|---|---|
| C-C | 2.41 | 3.4 |
| C-B | 3.29 | 3.4 |
| C-N | 4.07 | 3.37 |

In order to describe the diffusive motion of fullerene, we calculated mean square displacement (MSD) as following equation.

$$MSD = <\left(x(t)-x(0)\right)^2 + \left(y(t)-y(0)\right)^2> \quad (1)$$

Where, $x(t)$ and $y(t)$ are the coordinates of C60 centre of mass at the time $t$ and $<>$ represents averaging over the ensembles. In order to average over the ensembles, we divided the 8 ns simulation time, to 16 equal 500 ps time periods and then averaged the MSD over these 16 shorter intervals. In the case of 2D motions, mean square displacement is related to the time by Eq. (2).

$$MSD = 4Dt^{\alpha} \quad (2)$$

Where, $D$ and $\alpha$ are diffusion coefficient and diffusion anomaly parameter, respectively. In a typical diffusive motion, $\alpha = 1$. While when $\alpha > 1$, the motion regime is called super-diffusion and when $\alpha < 1$, the fullerene obeys sub-diffusion regime.

Since all of the substrates' atoms can move, it is possible for substrate to be displace during the simulation. Therefore, in all mentioned calculations, the relative motion of fullerene centre of mass and substrate centre of mass is assumed.

## Results and Discussion

Fig. 2 illustrates the equilibrium vertical distances between fullerene centre of mass and substrates. As it can be observed, C60 molecule has approximately equal distance to the substrate surfaces during the motion. For graphene and boron nitride substrates the average distances of C60 to the surfaces are 6.2 and 6.0 Å, respectively. Science graphene and boron nitride mono-layers have a same hexagonal structures, it is concluded that vertical equilibrium distance of fullerene is more affected by surface structure and less influenced by surface atom types.

Another parameter studied in this investigation was the Lennard-Jones potential energy between fullerene and surfaces. Boron nitride surface has shown greater interaction magnitude with C60 molecule than graphene layer. The last result is reasonable, since boron nitride has greater epsilon parameter in LJ potential than graphene, so boron nitride layer is expected to have stronger interaction with fullerene than graphene sheet. The average Lennard-Jones potential between fullerene and graphene has been calculated as -625.8 meV during the simulation time, while for boron nitride substrate the same quantity is -1024.5 meV.

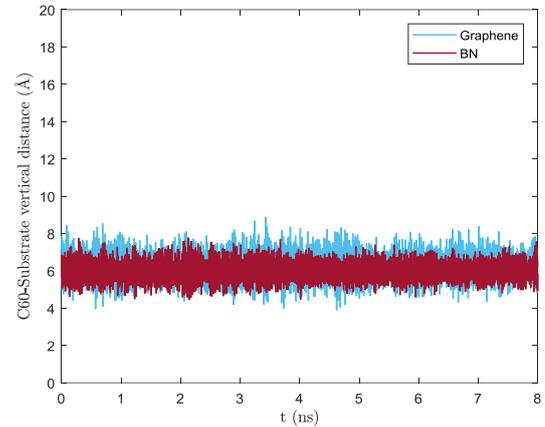

Fig. 2. C60 distance to the surfaces during simulation time

Fullerene trajectories on both surfaces are observable in Fig. 4. According to this figure, fullerene has more displacement range when it is moving on graphene surface. It can be as a result of the weaker interaction between C60 and graphene surface. The stronger interaction between boron nitride and fullerene causes the smaller displacement range on this surface.

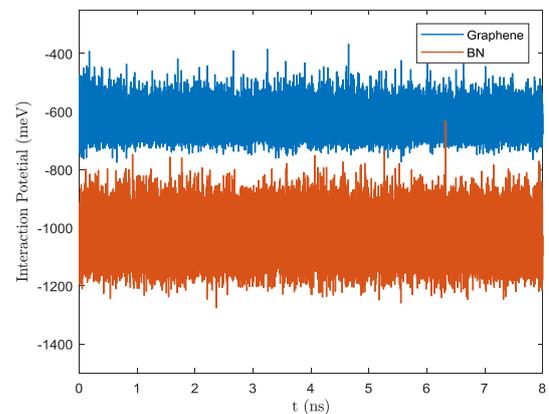

Fig. 3. C60-graphene and C60-boron nitride Lennard-Jones potentials during simulation time

In order to have a better perception of C60 diffusive motion, mean square displacement has been calculated. According to Eq. (2), we have a growth of MSD during the simulation time. In agreement with the larger displacements of C60 on graphene surface, here we observe that, C60 molecule has more MSD value on graphene surface in comparison with boron nitride surface at the same time. The difference of mean square displacements is $3.1470 \times 10^4 \text{Å}^2$ at the end of 500 ps. As a result, diffusion coefficient of the motion of fullerene on graphene surface is more than boron nitride substrate. If we approximately consider the motion of C60 as a normal diffusion on both substrates, diffusion coefficients are 28.05 and 15.08 $Å^2$/ps for graphene and boron nitride, respectively.

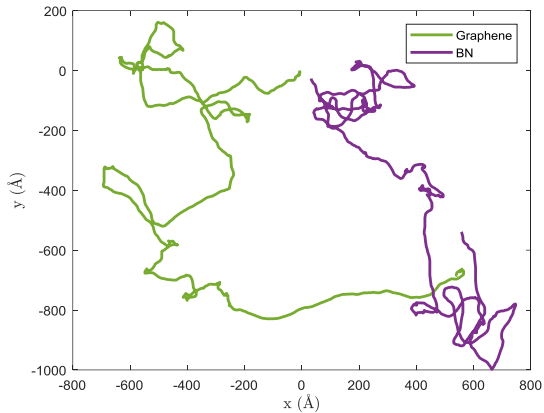

Fig. 4. Fullerene motion trajectories on graphene and boron nitride monolayers during the simulation time.

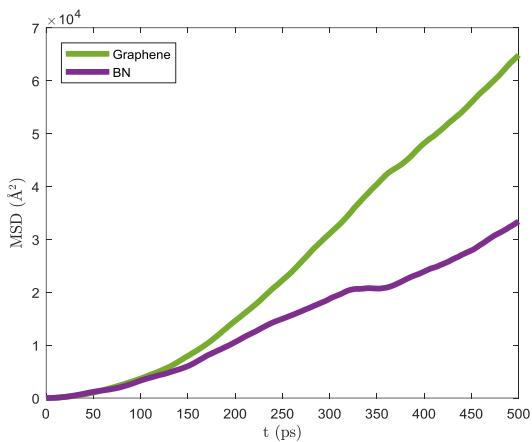

Fig. 5. Mean square displacement of the motion of fullerene on graphene and boron nitride monolayers.

Eventually, the logarithm of mean square displacement is studied as a function of time logarithm. According to Eq. (2), MSD diagram slop is proportional to diffusion anomaly parameter, while the intercept of the diagram illustrates the diffusion coefficient of motion. As it can be observed in Fig. 6, both diagrams have a same slope in first 50 ps and after that, a change in the diagrams slope is going to occur. Thus, at the beginning of the motion, fullerene has a same diffusion regime on graphene and boron nitride monolayers and after that a change in diffusion regime is observable. After first 50 ps, anomaly parameters of graphene and boron nitride are calculated as 1.828 and 1.469, respectively. As a result, C60 obeys a super-diffusion regime in both systems.

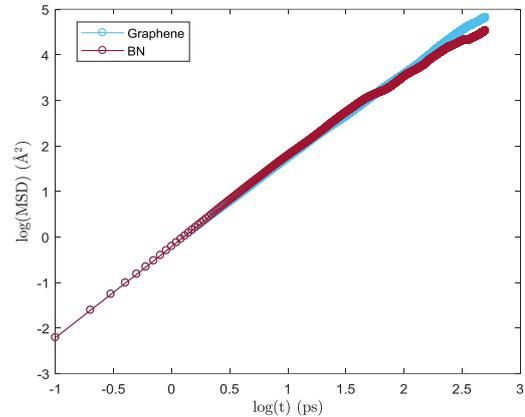

Fig. 6. Mean square displacement logarithm as a function of time logarithm for the motion of fullerene on graphene and boron nitride monolayers.

## Conclusions

We studied the motion of C60 molecule on 12nm×12nm graphene and boron nitride monolayers. For this purpose, classical molecular dynamic methods have been utilized. Both systems were studied at 300 K and with the same simulation conditions. Equilibrium distance of C60 molecule to the surface is approximately the same for both substrates. Lennard-Jones potential energy between fullerene and the surface illustrates a stronger interaction between C60 and boron nitride. The trajectory of C60 shows more displacement range, when fullerene is moving on the graphene surface. It has been shown that, fullerene has a greater diffusion coefficient on graphene surface. Finally, it is observed that, C60 motions on graphene and boron nitride obey a same super-diffusion regime at 300 K with different anomaly parameters.